\documentclass[final,3p,times,numbers]{elsarticle}

\biboptions{sort&compress}

\usepackage{epsfig}
\usepackage{amssymb}
\usepackage{amsmath}
\usepackage{subfigure}
\usepackage{dcolumn}
\usepackage{array} 
\usepackage{longtable} 

\usepackage[colorlinks]{hyperref}
\usepackage[usenames,dvipsnames]{color}
\hypersetup{
     breaklinks=true,
    pdfstartview={FitH},    
    colorlinks=true,       
    linkcolor=blue,          
    citecolor=red,        
    filecolor=magenta,      
    urlcolor=blue,           
    anchorcolor=green,      
    linktocpage=true
}
\usepackage{orcidlink}

\def\doi{http://doi.org}

\allowdisplaybreaks

\journal{Physics Letters B}

\begin{document}


\begin{frontmatter}

\title{Thermodynamic reconstruction in observationally constrained dynamical dark energy models} 

\author[1]{Sonej Alam\orcidlink{0009-0008-8322-2923}}
\ead{sonejalam36@gmail.com}

\affiliation[1]{organization={Department of Physics, Jamia Millia Islamia},
            city={New Delhi},
            postcode={110025}, 
            country={India}}

\begin{abstract}
We perform an observational and thermodynamic analysis of four cosmological models $\Lambda$CDM, CPL, MPL, and the three-parameter MmAH parametrization. All dynamical models improve the fit relative to $\Lambda$CDM, with $\Delta\chi^2 \sim 6$-$6.5$, with MPL providing the best fit, while MmAH remains viable.  Although Bayesian evidence mildly favors $\Lambda$CDM due to its lower complexity, the dynamical dark energy models remain competitive alternatives given current observations. We then reconstruct thermodynamic quantities within observationally constrained cosmological models, extending previous theoretical studies. The bestfit heat capacity reconstruction indicates that the divergence associated with a second order thermodynamic phase transition coincides with the deceleration-acceleration transition only in $\Lambda$CDM, whereas for the dynamical dark energy models it occurs at distinct redshifts, suggesting that this coincidence is not universal. The generalized second law is satisfied for all models over $0 \le z \le 1$, while the Hessian analysis reveals a transient instability at the phase transition; however, the late time thermodynamic stability is model dependent, with CPL and MPL remaining stable and $\Lambda$CDM and MmAH failing to satisfy both stability conditions simultaneously. These results show that thermodynamic properties reconstructed within observationally constrained cosmological models provide a complementary probe of dark energy, with the thermodynamic phase transition emerging as an intrinsically model dependent phenomenon.
\end{abstract}


\end{frontmatter}

\section{Introduction}

Over the past decade, a growing body of high precision cosmological observations has exposed significant tensions that challenge the internal consistency of the standard $\Lambda$ cold dark matter ($\Lambda$CDM) paradigm. The most striking among these is the $>5\sigma$ discrepancy between the locally measured value of the Hubble constant $H_0$ by the SH0ES collaboration~\cite{Riess:2021jrx} and the value inferred from cosmic microwave background (CMB) observations within the $\Lambda$CDM framework ~\cite{Planck:2018vyg}, commonly referred to as the Hubble tension. In addition, a persistent mismatch exists in the clustering amplitude parameter $S_8 = \sigma_8\sqrt{\Omega_{\rm m0}/0.3}$~\cite{Perivolaropoulos:2021jda,Kilo-DegreeSurvey:2023gfr}, where $\sigma_8$ quantifies the present day amplitude of matter fluctuations on scales of $8\,h^{-1}\,{\rm Mpc}$. These tensions may indicate the presence of unaccounted systematic effects or, more intriguingly, point toward new physics beyond the $\Lambda$CDM model~\cite{Kamionkowski:2022pkx,Freedman:2023jcz,Bernal:2016gxb,Knox:2019rjx}.

More recently, baryon acoustic oscillation (BAO) measurements from the Dark Energy Spectroscopic Instrument (DESI) have provided compelling indications in favor of dynamical dark energy (DDE). The first DESI data release (DR1) reported a $2.6\sigma$ preference for DDE over $\Lambda$CDM within the Chevallier–Polarski–Linder (CPL) parametrization~\cite{Chevallier:2000qy,Linder:2002et} when combined with CMB data~\cite{DESI:2024mwx}. This preference strengthens to $3.1\sigma$ in the second data release (DR2)~\cite{DESI:2025zgx,DESI:2025wyn}. Furthermore, incorporating various Type Ia supernova (SNeIa) datasets enhances the statistical significance to the $2.8\sigma$-$4.2\sigma$ range~\cite{DESI:2025zgx,DESI:2024aqx,DESI:2024kob,DESI:2025wyn}. Collectively, these developments have triggered extensive theoretical and phenomenological efforts to explore extensions of $\Lambda$CDM involving dynamical dark energy ~\cite{Gomez-Valent:2021cbe,Giare:2024smz,Berghaus:2024kra,Wolf:2025jlc,Lee:2025pzo,Zhong:2025gyn,Qu:2024lpx,Wang:2024dka,Giare:2024gpk,Gialamas:2024lyw,Shlivko:2024llw,Ye:2024ywg,Bhattacharya:2024hep,Ramadan:2024kmn,Jiang:2024xnu,Payeur:2024dnq,Malekjani:2024bgi,Wolf:2023uno,Wolf:2024eph,Wolf:2024stt,Chan-GyungPark:2024mlx,Park:2024vrw,Dinda:2024kjf,Dinda:2024ktd,Jiang:2024viw,Colgain:2024xqj,Bhattacharya:2024kxp,Akthar:2024tua,Chan-GyungPark:2025cri,Ferrari:2025egk,Peng:2025nez,Colgain:2024mtg,Mukherjee:2024ryz,Mukherjee:2025myk,Mukherjee:2025ytj,Berbig:2024aee,Li:2024qso,Li:2024qus,Li:2025owk,Borghetto:2025jrk,Carloni:2024zpl,Liu:2025mub,Wolf:2025jed,Chudaykin:2024gol,Sohail:2024oki,Hossain:2025grx,Cheng:2025lod,Scherer:2025esj,Pedrotti:2025ccw,Sohail:2025mma,Escamilla:2023oce,Liu:2025myr,Lu:2025gki,Wolf:2025acj,Cline:2025sbt,Wang:2025dtk,VanRaamsdonk:2025wvj,Peng:2025tqt,Yang:2025oax,Jiang:2025hco,Ishak:2025cay,Shlivko:2025fgv,Silva:2025twg,Gialamas:2025pwv,Cortes:2025joz,Artola:2025zzb,Dinda:2025hiu,Park:2025fbl,Blanco:2025vva,Li:2025muv,Zhang:2025dwu,Alam:2025epg,Hossain:2025gpr,Nojiri:2025low,Nojiri:2025uew,2025IJMPD..3450058P}.

A widely used framework for probing dynamical dark energy is the two parameter CPL parametrization $w(a)=w_0+w_a(1-a)$, which corresponds to a first-order Taylor expansion of the dark energy equation of state around the present scale factor $a=1$. While CPL is economical and analytically tractable, it may be too restrictive to capture possible high redshift transitions. Extensions involving higher order terms introduce additional freedom but often lead to poorly constrained parameters~\cite{Nesseris:2025lke}. To overcome these limitations, the Akthar–Hossain (AH) parametrization~\cite{Akthar:2024tua} was proposed as a general framework capable of describing thawing, freezing, and tracker behaviours. Its minimal version (mAH) enforces $w \rightarrow -1$ at high redshift, which restricts its dynamical freedom and leads to weaker statistical performance relative to CPL or $w$CDM~\cite{Akthar:2024tua}. To address this limitation, we recently proposed an extension—the modified minimal AH (MmAH) parametrization—by promoting the high redshift asymptote of the equation of state to a free parameter~\cite{Alam:2025epg}. This introduces an additional degree of freedom, allowing for a broader class of dark energy dynamics.


Beyond statistical comparisons, thermodynamic considerations provide an independent and physically motivated probe of cosmological models. The concept of horizon thermodynamics, inspired by black hole physics, has been widely used to test the validity of cosmological scenarios through the generalized second law (GSL), which states that the total entropy of the universe never decreases~\cite{bekenstein1974generalized}. For a comprehensive discussion, see~\cite{faraoni2015horizons}. The thermodynamic stability of a system can be inferred from the second derivatives of its entropy. For example, Barboza et al.~\cite{Barboza:2015bha} argued that stability requires positive heat capacities, leading to tension with observations, while Luongo and Quevedo~\cite{luongo2014cosmographic} showed that an accelerating universe is naturally associated with a negative heat capacity at constant volume, consistent with $\Lambda$CDM. More recently, Duary, Banerjee and Dasgupta~\cite{Duary:2023nnf} demonstrated that in a $\Lambda$CDM-like model the heat capacity exhibits a discontinuity at the redshift where the universe transitions from deceleration to acceleration, signaling a second-order phase transition with the deceleration parameter acting as an order parameter. This result crucially depends on the use of the Hayward–Kodama temperature. In a subsequent study~\cite{Duary:2019dfu}, they found that certain quintessence models violate the GSL.

In this work, we extend these thermodynamic analyses to observationally constrained dark energy models. For each model, we reconstruct the deceleration parameter $q(z)$, heat capacity $C_V(z)$, entropy production rate $\dot{S}_{\text{tot}}(z)$, and Hessian stability minors $S_{UU}(z)$ and $\alpha(z)$ directly from the MCMC chains. This allows us to test whether the thermodynamic phase transition at $q=0$ is a universal feature or model dependent, and whether the generalized second law holds in realistic cosmological scenarios. The primary objective of this work is therefore twofold: first, to perform a consistent and unified observational analysis of $\Lambda$CDM, CPL, MPL, and MmAH using current cosmological data, and second, to investigate their thermodynamic properties in a data driven framework. Parameter estimation is carried out using the nested sampling algorithm PolyChord, which is well suited for efficiently exploring the degenerate parameter spaces characteristic of dynamical dark energy models.

The paper is organised as follows. Section~\ref{sec:thermo} presents the thermodynamic formalism. Section~\ref{sec:models} introduces the cosmological models. Section~\ref{sec:data} describes the data and methodology. Section~\ref{sec:result} presents the results, and Section~\ref{sec:con} concludes.

\section{Thermodynamic Formalism}
\label{sec:thermo}

We consider a spatially flat Friedmann–Robertson–Walker (FRW) universe with metric
\begin{equation}
ds^2 = -dt^2 + a^2(t)\left[dr^2 + r^2 d\Omega^2\right],
\end{equation}
where $a(t)$ is the scale factor. The Hubble parameter is $H = \dot{a}/a$, and an overdot denotes a derivative with respect to cosmic time $t$. The Friedmann equations for a perfect fluid of total energy density $\rho$ and pressure $p$ are
\begin{align}
3H^2 &= \rho, \label{eq:friedmann} \\
2\dot{H} &= -(\rho + p), \label{eq:accel}
\end{align}
where we use units $8\pi G = c = 1$.

\subsection{Apparent horizon and its thermodynamics}

For a spatially flat FRW universe, the apparent horizon is located at the comoving radius $\tilde{r}_h = 1/H$ \cite{Bak:1999hd,Ferreira:2015iaa,Hawking:1975vcx,Faraoni:2015ula}. Its area is $A = 4\pi\tilde{r}_h^2$, and, by analogy with black hole thermodynamics, its entropy is one quarter of the area (in Planck units):
\begin{equation}
S_h = \frac{A}{4G} = 8\pi^2 \tilde{r}_h^2 = \frac{8\pi^2}{H^2}. \label{eq:horizon_entropy}
\end{equation}
The temperature of a dynamical horizon is defined via the Hayward–Kodama surface gravity \cite{hayward1998unified,Kodama:1979vn}. For the FRW metric, the surface gravity is $\kappa_{\text{ko}} = -(2H^2+\dot{H})/(2H)$ \cite{Cai}, and the corresponding Hayward–Kodama temperature \cite{hayward2009local,DiCriscienzo:2009kun} is
\begin{equation}
T_h = \frac{|\kappa_{\text{ko}}|}{2\pi} = \frac{2H^2 + \dot{H}}{4\pi H}. \label{eq:temp}
\end{equation}
In a de Sitter universe ($\dot{H}=0$) this reduces to the Gibbons–Hawking temperature $T = H/(2\pi)$ \cite{Hawking:1974rv}.

\subsection{Generalized second law}

We assume that the fluid inside the apparent horizon is in thermal equilibrium with the horizon. The first law of thermodynamics applied to the fluid gives
\begin{equation}
T_h dS_{\text{in}} = dU + p dV, \label{eq:first_law}
\end{equation}
where $S_{\text{in}}$ is the entropy of the fluid, $U = \rho V$ its internal energy, and $V = \frac{4}{3}\pi\tilde{r}_h^3 = \frac{4\pi}{3H^3}$ the volume inside the horizon. Taking the time derivative and using the continuity equation $\dot{\rho} = -3H(\rho+p)$ together with $\dot{\tilde{r}}_h = -\dot{H}/H^2$, we obtain
\begin{align}
\dot{S}_{\text{in}} &= \frac{1}{T_h}\left[(\rho+p)\dot{V} + \dot{\rho}V\right] \nonumber\\
&= 16\pi^2 \frac{\dot{H}}{H^3}\left(1 + \frac{\dot{H}}{2H^2+\dot{H}}\right). \label{eq:Sdot_in}
\end{align}
The horizon entropy changes at a rate
\begin{equation}
\dot{S}_h = \frac{d}{dt}\left(\frac{8\pi^2}{H^2}\right) = -16\pi^2 \frac{\dot{H}}{H^3}. \label{eq:Sdot_h}
\end{equation}
Adding the two contributions yields the total entropy production rate
\begin{equation}
\dot{S}_{\text{tot}} = \dot{S}_h + \dot{S}_{\text{in}} = 16\pi^2 \frac{\dot{H}^2}{H^3}\left(\frac{1}{2H^2+\dot{H}}\right). \label{eq:Sdot_tot}
\end{equation}
The generalized second law (GSL) requires $\dot{S}_{\text{tot}} \ge 0$. Since $\dot{H}^2 \ge 0$, the sign of $\dot{S}_{\text{tot}}$ is determined solely by the denominator $2H^2+\dot{H}$. When this denominator becomes negative, the GSL is violated in the original formalism.

\subsection{Heat capacity and phase transition}

The heat capacity at constant volume is defined as $C_V = T(\partial S_{\text{in}}/\partial T)_V$. Using the first law and the expressions for $H$ and $\dot{H}$, one derives \cite{Duary:2023nnf}
\begin{equation}
C_V = 32\pi^2 \frac{\dot{H}}{2H^2\dot{H} + H\ddot{H} - \dot{H}^2}. \label{eq:CV}
\end{equation}
A divergence of $C_V$ occurs when the denominator vanishes,
\begin{equation}
2H^2\dot{H} + H\ddot{H} - \dot{H}^2 = 0. \label{eq:CVdiv}
\end{equation}
Such a divergence signals a second‑order phase transition: the heat capacity becomes infinite while the entropy itself remains continuous. In the $\Lambda$CDM‑mimicking model of Ref.~\cite{Duary:2023nnf}, this divergence coincides with the deceleration parameter $q = -1-\dot{H}/H^2$ crossing zero, and $q$ acts as an order parameter.

\subsection{Thermodynamic stability}

Local thermodynamic stability is assessed by examining the Hessian matrix of the entropy $S_{\text{in}}$ with respect to the internal energy $U$ and volume $V$ \cite{callen1998thermodynamics,kubo1968thermodynamics,carter2000classical,Muller1985}:
\begin{equation}
W = \begin{bmatrix}
\dfrac{\partial^2 S_{\text{in}}}{\partial U^2} & \dfrac{\partial^2 S_{\text{in}}}{\partial U\partial V} \\[6pt]
\dfrac{\partial^2 S_{\text{in}}}{\partial V\partial U} & \dfrac{\partial^2 S_{\text{in}}}{\partial V^2}
\end{bmatrix}.
\end{equation}
For the entropy to be maximized (i.e., for thermodynamic stability), all principal minors of $W$ must satisfy: the first‑order minor $\partial^2 S_{\text{in}}/\partial U^2 \le 0$, and the determinant $\alpha \equiv \det W \ge 0$. Using the first law, these conditions become
\begin{align}
\frac{\partial^2 S_{\text{in}}}{\partial U^2} &= -\frac{1}{T^2 C_V} \le 0, \label{eq:SUU}\\
\alpha &= \frac{1}{C_V T^3 V \beta_T} \ge 0, \label{eq:alpha}
\end{align}
where $\beta_T = -\frac{1}{V}\left(\frac{\partial V}{\partial p}\right)_T$ is the isothermal compressibility. The isothermal compressibility can be expressed in terms of $H$ and its derivatives as
\begin{equation}
\beta_T = \frac{3\dot{H}(2H^2+\dot{H})}{2(H^2+\dot{H})(2H^2\dot{H}+H\ddot{H}-\dot{H}^2)}. \label{eq:betaT}
\end{equation}
Because $C_V$ appears in the denominator of both $S_{UU}$ and $\alpha$, a negative heat capacity (which is common in gravitational systems \cite{padmanabhan1990statistical,luongo2014cosmographic}) automatically violates the first stability condition ($S_{UU}>0$), indicating that the fluid is not thermodynamically stable. The GSL condition $\dot{S}_{\text{tot}} \ge 0$ is a separate, weaker requirement that must be satisfied even for unstable systems.

In summary, all thermodynamic quantities of interest – $q(z)$, $T_h(z)$, $\dot{S}_{\text{tot}}(z)$, $C_V(z)$, $S_{UU}(z)$, and $\alpha(z)$ – can be computed once the Hubble parameter $H(z)$ and its derivatives $\dot{H}(z)$ and $\ddot{H}(z)$ are known for a given dark energy model. In the following sections we apply this formalism to several dynamical dark energy models constrained by current cosmological data.

\section{Dynamical Dark Energy Models}
\label{sec:models}

Besides the standard $\Lambda$CDM scenario and the CPL parametrization, we explore alternative dynamical dark energy extensions that allow additional freedom in the redshift evolution of the equation of state, particularly beyond the very low redshift regime.


\subsection{Modified Power-Law (MPL) Parametrization}

To ensure a well behaved evolution at high redshift while maintaining sufficient flexibility at late times, we consider the modified power-law (MPL) parametrization \cite{Hossain:2025gpr},
\begin{equation}
    w(a) = \frac{2 w_0}{1 + a^{\gamma}} \, ,
    \label{eq:mpl_eos}
\end{equation}
which satisfies $w(a=1)=w_0$ and remains finite in the limit $a \to 0$. In the early-time limit, this parametrization approaches a constant value $w \to 2w_0$, ensuring a regular asymptotic behavior.

Expanding around the present epoch ($a=1$), the effective CPL slope is given by
\begin{equation}
    w_a = -\left.\frac{dw}{da}\right|_{a=1} = \frac{\gamma w_0}{2} \, .
\end{equation}

The corresponding evolution of the dark energy density is
\begin{equation}
    \rho_{\rm DE}(z) = \rho_{\rm DE,0}(1+z)^3
    \left(\frac{1 + (1+z)^{\gamma}}{2}\right)^{\frac{6 w_0}{\gamma}},
    \label{eq:rho_mpl_new}
\end{equation}
which smoothly reduces to the $w$CDM limit as $\gamma \to 0$. At high redshift, the asymptotic scaling behaves as $\rho_{\rm DE} \propto (1+z)^{3+6w_0}$.

\subsection{Modified mAH Parametrization}

To further enhance the flexibility in the redshift evolution of the dark energy equation of state, we consider a modified version of the mAH parametrization\cite{Alam:2025epg},
\begin{equation}
     w(z) = \beta + \frac{\gamma z}{1 + \dfrac{1+z}{\Omega_\delta}} \, ,
    \label{eq:wz-main}
\end{equation}
where $\gamma$, $\beta$, and $\Omega_\delta$ are free parameters controlling the amplitude, baseline value, and transition scale, respectively.

This form can be viewed as a specific case of a more general parametrization,
\begin{equation}
    w(z) = \beta + \frac{\gamma \, z^\xi}{1 + \left(\dfrac{1+z}{\Omega_\delta}\right)^\zeta},
\end{equation}
for suitable choices of the constants $\xi$ and $\zeta$.

For the parametrization~\eqref{eq:wz-main}, the asymptotic limits are
\begin{align}
    w(z\to \infty) &= \beta+\gamma\Omega_\delta, \\
    w(z\to -1) &= \beta - \gamma.
\end{align}

At the transition scale defined by $1+z_t=\Omega_\delta$, the equation of state becomes
\begin{equation}
    w(z_t) = \beta + \frac{\gamma z_t}{2}.
\end{equation}

Expanding around the present epoch ($z=0$), the effective CPL parameters are
\begin{align}
    w_0 &= \beta, \\
    w_a &= \frac{\gamma\Omega_\delta}{1+\Omega_\delta}, \\
    w_b &= \frac{\gamma \, \Omega_\delta^2}{(1+\Omega_\delta)^2}.
\end{align}

This parametrization allows both positive and negative values of $w_a$, thereby accommodating a wider range of possible dark energy dynamics compared to more restrictive models.

The evolution of the dark energy density is given by
\begin{equation}
    \rho_{\rm DE}(z) = \rho_{\rm DE,0}
    (1+z)^{3(1+\beta-\gamma)}
    \left(
      \frac{\Omega_\delta+1+z}{\Omega_\delta+1}
    \right)^{3\gamma(1+\Omega_\delta)}.
    \label{eq:rhoDE_MmAH}
\end{equation}

\section{Observational Data and Methodology}
\label{sec:data}

To constrain the cosmological models, we perform a joint likelihood analysis combining multiple cosmological probes. The analysis is carried out within a Bayesian framework using the Cobaya package~\citep{Torrado:2020dgo}, which provides a flexible interface for likelihood evaluation and parameter estimation. The theoretical predictions for both the background and perturbation evolution are computed using the Boltzmann solver CLASS ~\cite{2011JCAP...07..034B}. The priors adopted for the cosmological and dark energy parameters are summarized in Table~\ref{priors}.

\begin{table}[ht]
\centering
\caption{Uniform priors adopted for the sampled cosmological and dark-energy parameters.}
\label{priors}
\begin{tabular}{lc}
\hline
Parameter & Prior Range \\
\hline
$\log(10^{10}A_s)$ & $[1.61,;3.91]$ \\
$n_s$ & $[0.8,;1.2]$ \\
$H_0$ & $[50,;90]$ \\
$\omega_b \equiv\Omega_b h^2$ & $[0.005,;0.1]$ \\
$\omega_{cdm}\equiv\Omega_{\rm c} h^2$ & $[0.001,;0.99]$ \\
\hline
$w_0$ & $[-2,;0.]$ \\
$w_a$ & $[-2.,;1.0]$ \\
\hline
$\gamma {(MPL)}$ & $[0,;5]$ \\
\hline
$\gamma{(MmAH)}$ & $[-1.5,;1.0]$ \\
$\Omega_{\delta}$ & $[0,;100]$ \\
\hline
\end{tabular}
\end{table}

Parameter estimation is performed using the nested sampling algorithm PolyChord~\cite{Handley:2015fda}, which is particularly effective for exploring degenerate and multi-dimensional parameter spaces. In addition to obtaining posterior distributions, PolyChord computes the Bayesian evidence
\begin{equation}
\ln Z=\ln \int \mathcal{L}(D|\theta,M)P(\theta|M)d\theta,
\end{equation}

where $\mathcal{L}(D|\theta,M)$ is the likelihood, $P(\theta|M)$ is the prior probability distribution, and $\theta$ denotes the set of sampled parameters. The Bayesian evidence is subsequently used for model comparison.

The bestfit parameters are determined via maximum likelihood estimation, and the goodness-of-fit is quantified through
\begin{equation}
\chi^2_{\rm min} = -2 \ln \mathcal{L}_{\rm max}.
\end{equation}

\subsection{Cosmic Microwave Background}

We use temperature and polarization anisotropies of the cosmic microwave background (CMB), along with lensing information. The likelihood includes the high-multipole ($\ell > 30$) power spectra $C_{\ell}^{TT}$, $C_{\ell}^{TE}$, and $C_{\ell}^{EE}$ from the PR4 \texttt{CamSpec} likelihood~\cite{refId0}. In addition, we incorporate CMB lensing measurements from the PR4 NPIPE reconstruction~\cite{Carron:2022eyg}. The combined dataset is hereafter referred to as the CMB dataset.

\subsection{Baryon Acoustic Oscillations}

The BAO measurements are taken from DESI DR2~\cite{DESI:2025zgx}, comprising more than 14 million extragalactic objects, including emission line galaxies (ELGs), luminous red galaxies (LRGs), quasars (QSOs)~\cite{DESI:2025qqy}, and Ly$\alpha$ forest tracers~\cite{DESI:2025zpo}. These observations provide constraints on the distance ratios $D_{M}/r_{d}$ and $D_{H}/r_{d}$ over the redshift range $0.4 < z < 4.2$, along with low-redshift measurements of $D_{V}/r_{d}$ in the range $0.1 < z < 0.4$, incorporating both anisotropic and isotropic BAO analyses.

\subsection{Type Ia Supernovae}

For Type Ia supernovae, we use the Pantheon$+$ compilation, which consists of 1550 luminosity distance measurements spanning the redshift range $0.001 < z < 2.26$~\cite{Scolnic:2021amr,Brout:2022vxf}.

\section{Results}
\label{sec:result}
\begin{figure*}[ht]
\centering
\includegraphics[scale=0.25]{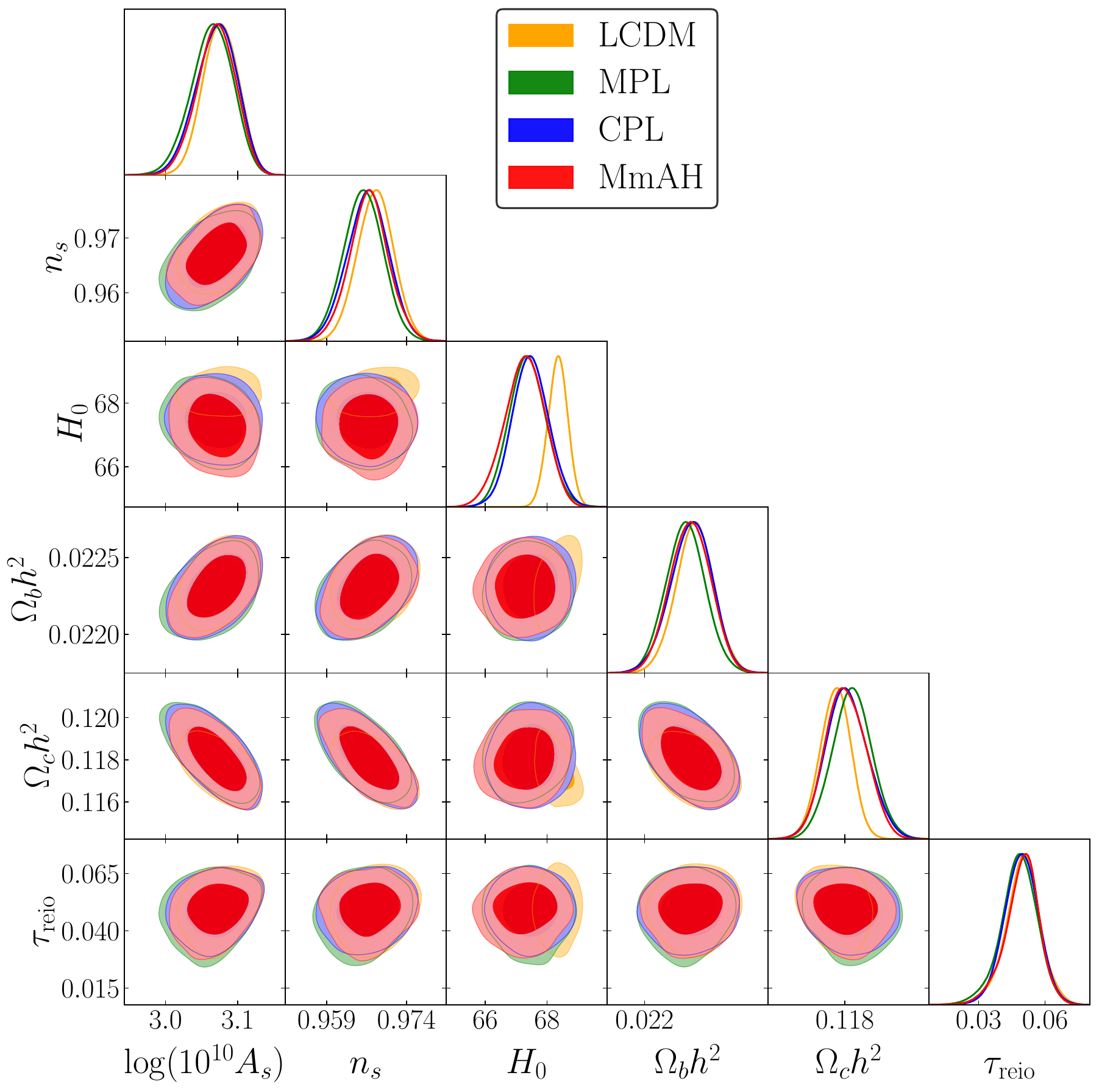}
\includegraphics[scale=0.4]{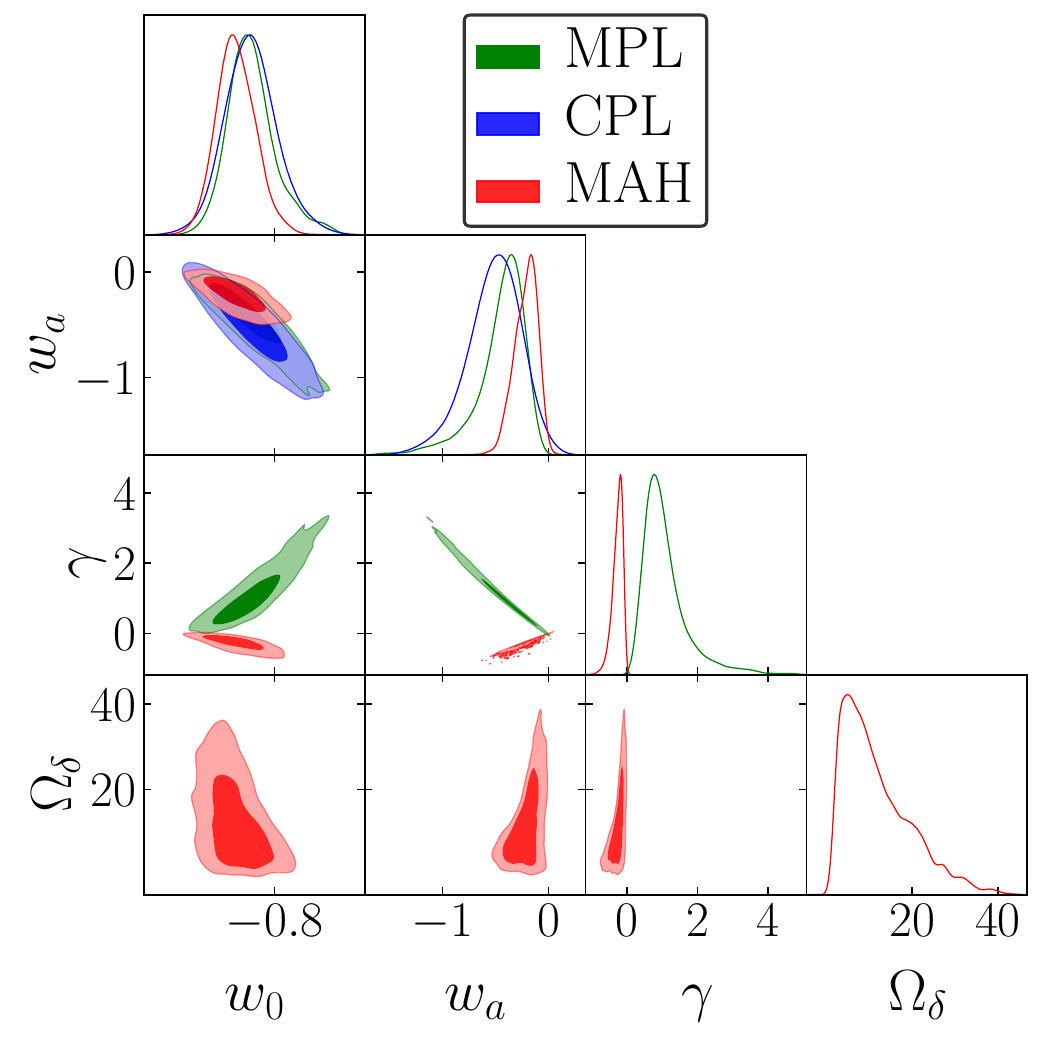}
\caption{Posterior distributions of cosmological parameters from the combined \textit{Planck+DESI DR2 BAO+Pantheon+} data.}
\label{fig:contours}
\end{figure*}

We present the cosmological constraints obtained from the combined Planck + DESI DR2 BAO + Pantheon+ datasets. The marginalized parameter constraints for $\Lambda$CDM, CPL, MPL, and MmAH models are summarized in Table~\ref{tab:constraints_combined}, while the corresponding posterior distributions are shown in Fig.~\ref{fig:contours}.

\begin{table*}[t]
\begin{center} 
\caption{Constraints on the parameters and model comparison statistics for $\Lambda$CDM, CPL, MPL and MmAH models using the Planck + DESI DR2 BAO + Pantheon+ datasets. All parameter constraints are quoted at the $68\%$ confidence level ($1\sigma$), and upper or lower limits correspond to the same confidence level.}
\label{tab:constraints_combined} 
\resizebox{\textwidth}{!}{%
\begin{tabular}{lcccc} 
\hline\hline 
\multicolumn{5}{c}{\textbf{Planck + DESI DR2 BAO + Pantheon+}}\\ 
\hline 
Parameter & $\Lambda$CDM & CPL &  MPL & MmAH \\ 
\hline 
$100\,\omega_{\rm b}$ & $2.23(2.235)\pm 0.017$ & $2.2314(2.231) \pm 0.081$  & $2.2276(2.227) \pm 0.016$ & $2.23(2.232) \pm 0.019$ \\ 
$\omega_{\rm cdm}$ & $0.11754(0.11746)^{+0.00099}_{-0.00087}$ & $0.1181(0.1181) \pm 0.0013$  & $0.118359(0.1185) \pm 0.0011$ & $0.11805(0.1180)\pm 0.0013$ \\ 
$100\,\theta_s$ & $1.04189(1.04191)\pm 0.00033$ & $1.04184(1.04185) \pm 0.00032$ & $1.04183(1.04183) \pm 0.00029$ & $1.04185(1.04186)\pm 0.00032$ \\ 
$\ln(10^{10}A_s)$ & $3.07642(3.077)\pm 0.03$ & $3.07047(3.07) ^{+0.035}_{-0.030}$  & $3.06431(3.063)^{+0.033}_{-0.029}$ & $3.07028(3.072) \pm 0.032$\\ 
$n_s$ & $0.968(0.9685)\pm 0.0045$ & $0.96674(0.9668) \pm 0.0048$  & $0.965878(0.9659) \pm 0.0043$ & $0.967(0.9675)^{+0.0043}_{-0.0049} $ \\ 
$\tau_{\rm reio}$ & $0.0498961(0.05)^{+0.0091}_{-0.011}$ & $0.0492304(0.050) \pm 0.011$  & $ 0.048134(0.049)\pm 0.010$ & $0.0495241(0.0521)^{+0.008}_{-0.011}$ \\
$w_0$ & \dots & $-0.856611(-0.85)^{+0.060}_{-0.074}$  & $-0.85233(-0.843)^{+0.044}_{-0.076}$ & $-0.888514(-0.887)^{+0.047}_{-0.054}$ \\ 
$w_a$ & \dots & $-0.495535(-0.54)^{+0.33}_{-0.23}$  & $-0.52^{+0.2}_{-0.19}$ & $ -0.21 ^{+0.09}_{-0.12} $ \\
$\gamma$ & \dots & \dots  & $1.06698(1.24)^{+0.22}_{-0.90}$ & $-0.252516(-0.279)^{+0.21}_{-0.095}$ \\
$\Omega_\delta$ & \dots & \dots & \dots  & $<11$ \\ 
$z_{\rm t}$ & \dots & \dots & \dots   & $<10$ \\ 
\hline
$H_0$ &$68.35(68.31)^{+0.44}_{-0.36}$ & $67.46(67.47)\pm 0.77$  & $67.37(67.39)\pm 0.72$ &$67.275(67.26)^{+0.84}_{-0.73}$ \\[1ex] 
$\sigma_8$ &$0.8198(0.817)^{+0.013}_{-0.010}$ & $0.812044(0.81)^{+0.015}_{-0.011}$  & $0.809(0.809)^{+0.014}_{-0.010}$ & $0.81(0.809)^{+0.014}_{-0.010}$ \\[1ex]
\hline 
$\chi^2_{\rm min}$ & $12390.909$ & $12384.75$ & $12384.371$ & $12384.564$ \\ 
$\Delta\chi^2_{\rm min}$ & $0$ & $-6.159$  & $-6.538$ & $-6.345$ \\
$\ln \mathcal{Z}$ & $-6244.4$ & $-6247.15$ & $-6247.22$ & $-6249.26$ \\ 
$\Delta$$\ln \mathcal{Z}$ & $0$ & $-2.75$  & $-2.82$ & $-4.86$ \\ 
\hline 
\end{tabular}} 
\end{center} 
\end{table*}

We first note that the standard cosmological parameters remain remarkably stable across all models. In particular, the baryon density $100\,\omega_{\rm b}$, cold dark matter density $\omega_{\rm cdm}$, and the angular acoustic scale $100\,\theta_s$ are consistent within $1\sigma$ for all parametrizations considered. Similarly, the scalar spectral index $n_s$, amplitude $\ln(10^{10}A_s)$, and optical depth $\tau_{\rm reio}$ show no statistically significant deviations from their $\Lambda$CDM values. This stability indicates that the extensions explored here primarily modify the late time expansion history, while leaving the early Universe physics, tightly constrained by CMB observations, essentially unchanged.

The constraints on the dark energy sector reveal clear signatures of dynamical behaviour. For the CPL parametrization, we obtain $w_0 = -0.85^{+0.06}_{-0.074}$ and $w_a = -0.54^{+0.33}_{-0.23}$, indicating a time evolving equation of state. The MPL model yields $w_0 = -0.843^{+0.044}_{-0.076}$ with $\gamma = 1.24^{+0.22}_{-0.90}$, while the MmAH parametrization gives $w_0 = -0.887^{+0.047}_{-0.054}$ and $\gamma = -0.279^{+0.21}_{-0.095}$, along with upper bounds on the additional parameter $\Omega_\delta$. A common feature across all dynamical models is that $w_0 < -1$, suggesting a mild but consistent deviation from a pure cosmological constant. From a frequentist perspective, all dynamical dark energy models provide an improved fit relative to $\Lambda$CDM. The reductions in the minimum $\chi^2$ are $\Delta\chi^2 = -6.159$ for CPL, $-6.538$ for MPL, and $-6.355$ for MmAH.When Bayesian model comparison is considered, however, a different picture emerges. The Bayesian evidence values reported in Table~\ref{tab:constraints_combined} indicate that $\Lambda$CDM remains mildly preferred over the dynamical dark energy parametrizations, with $\Delta\ln\mathcal{Z}=-2.75$, $-2.82$, and $-4.86$ for CPL, MPL, and MmAH, respectively. This reflects the Occam penalty associated with the additional model parameters. Nevertheless, the relatively small evidence differences for CPL and MPL suggest that these parametrizations remain competitive alternatives to $\Lambda$CDM given current observations. Interestingly, a similar conclusion has recently been reported in \cite{Ong:2026tta}, where Bayesian evidence was also found to mildly favor $\Lambda$CDM despite indications of dynamical dark energy from parameter estimation analyses.

Among the dynamical dark energy models, the MPL parametrization provides the largest improvement in $\chi^2$, followed closely by MmAH. Although this improvement does not translate into a preference in Bayesian evidence, it indicates that the additional degree of freedom encoded in $\gamma$ captures features of the late time expansion history that are not accounted for in the standard $\Lambda$CDM model. Future high precision observations will be required to determine whether these improvements reflect genuine physical effects or merely the increased flexibility of the parametrizations.

\subsection{Deceleration parameter and phase transition}

\begin{figure*}[ht]
\centering
\includegraphics[scale=0.37]{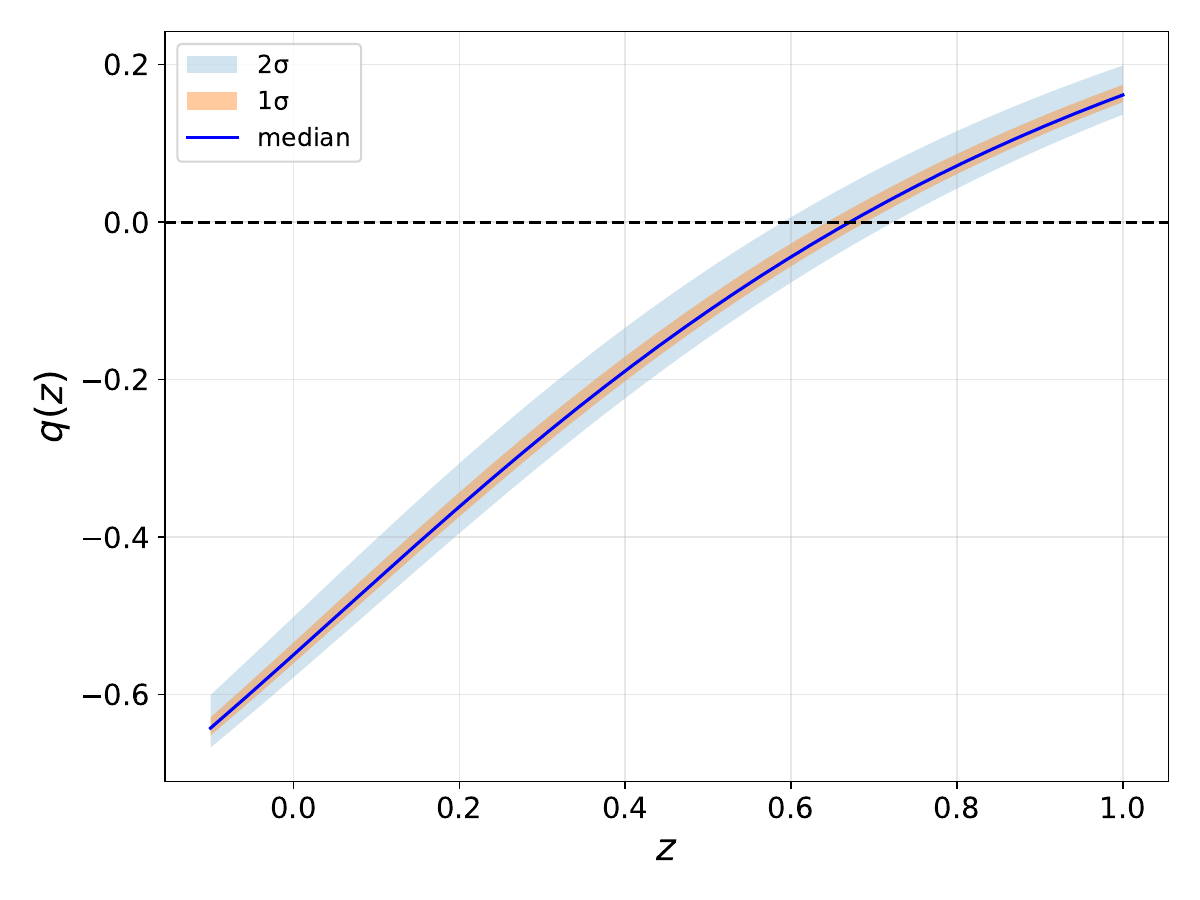}
\includegraphics[scale=0.37]{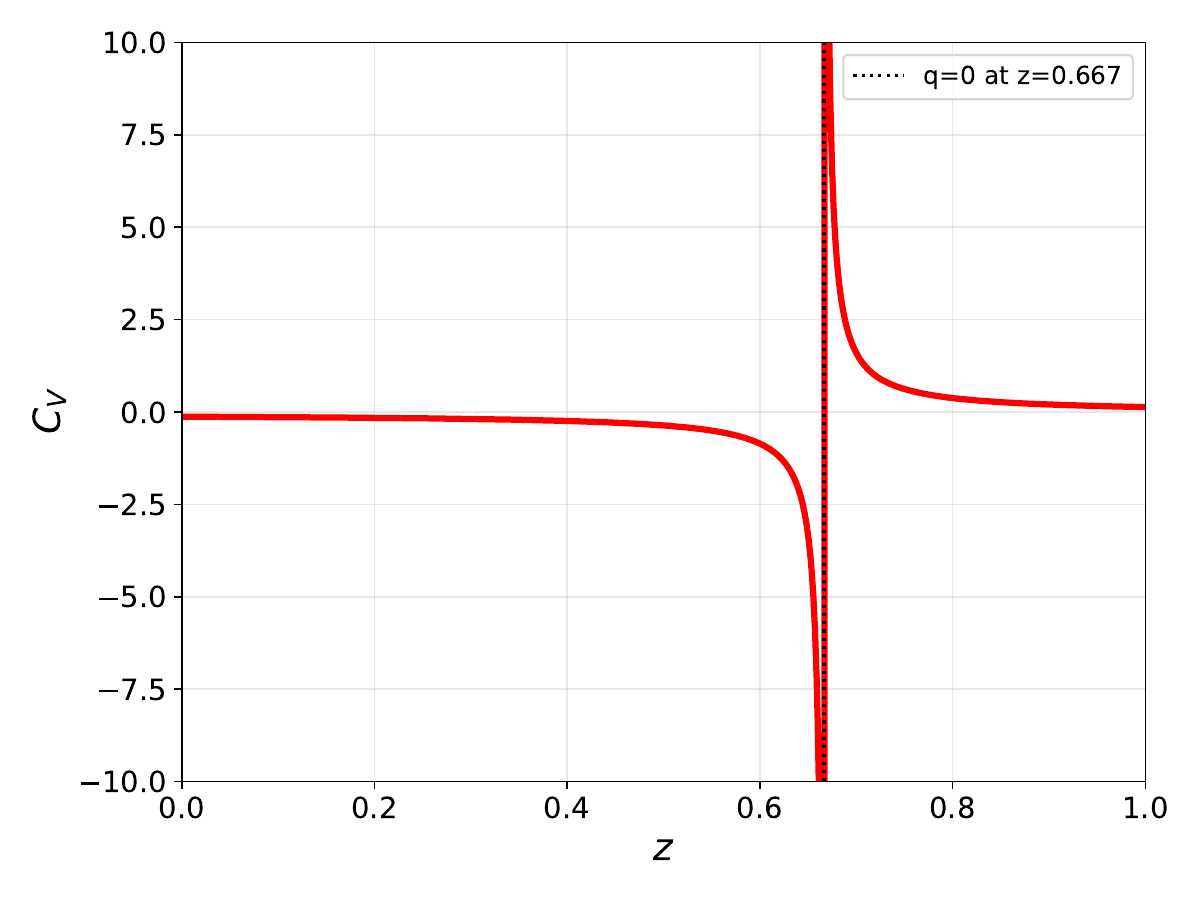}
\caption{$\Lambda$CDM: Deceleration parameter $q(z)$ (left) and best‑fit heat capacity $C_V(z)$ (right). The vertical dashed line marks $z_{q=0}$.}
\label{fig:q_cv_LCDM}
\end{figure*}

\begin{figure*}[ht]
\centering
\includegraphics[scale=0.37]{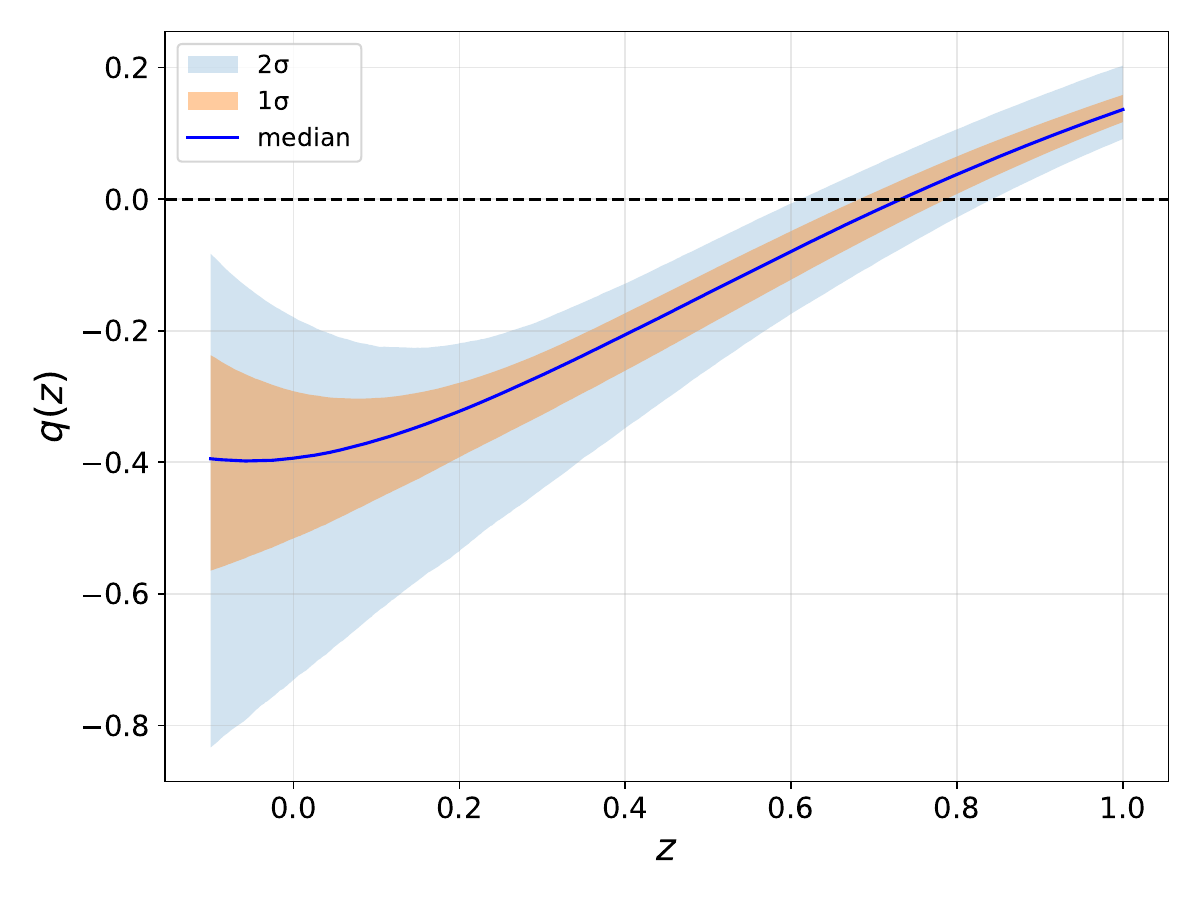}
\includegraphics[scale=0.37]{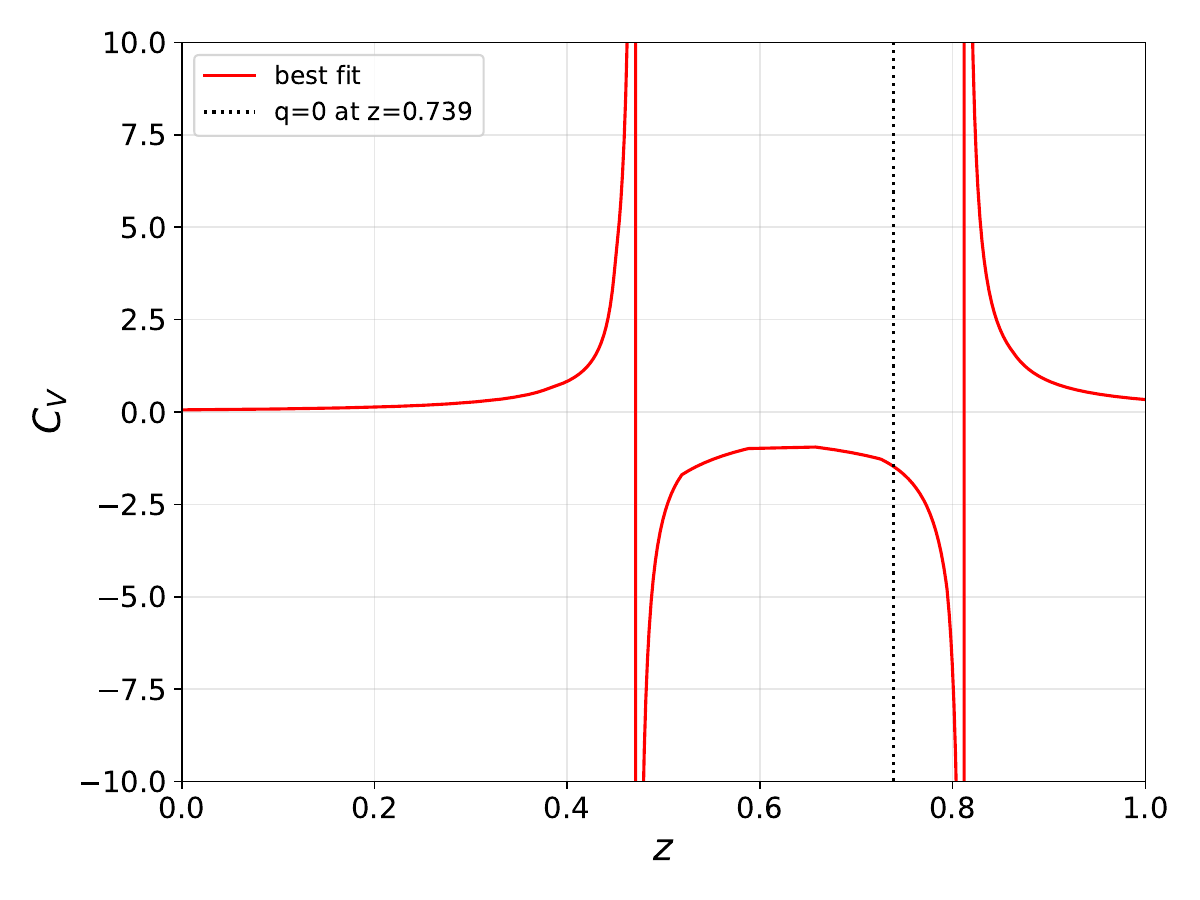}
\caption{CPL model: same as Fig.~\ref{fig:q_cv_LCDM}.}
\label{fig:q_cv_CPL}
\end{figure*}

\begin{figure*}[ht]
\centering
\includegraphics[scale=0.37]{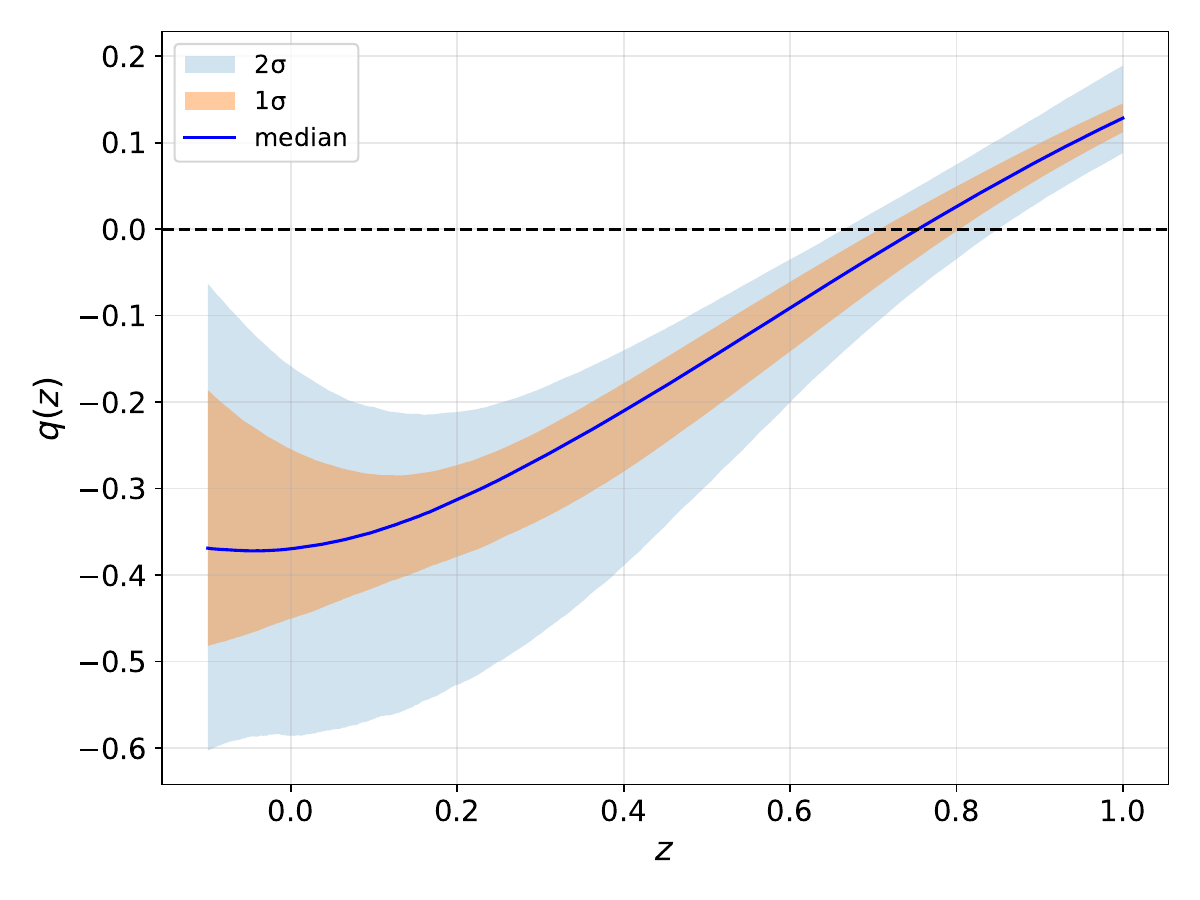}
\includegraphics[scale=0.37]{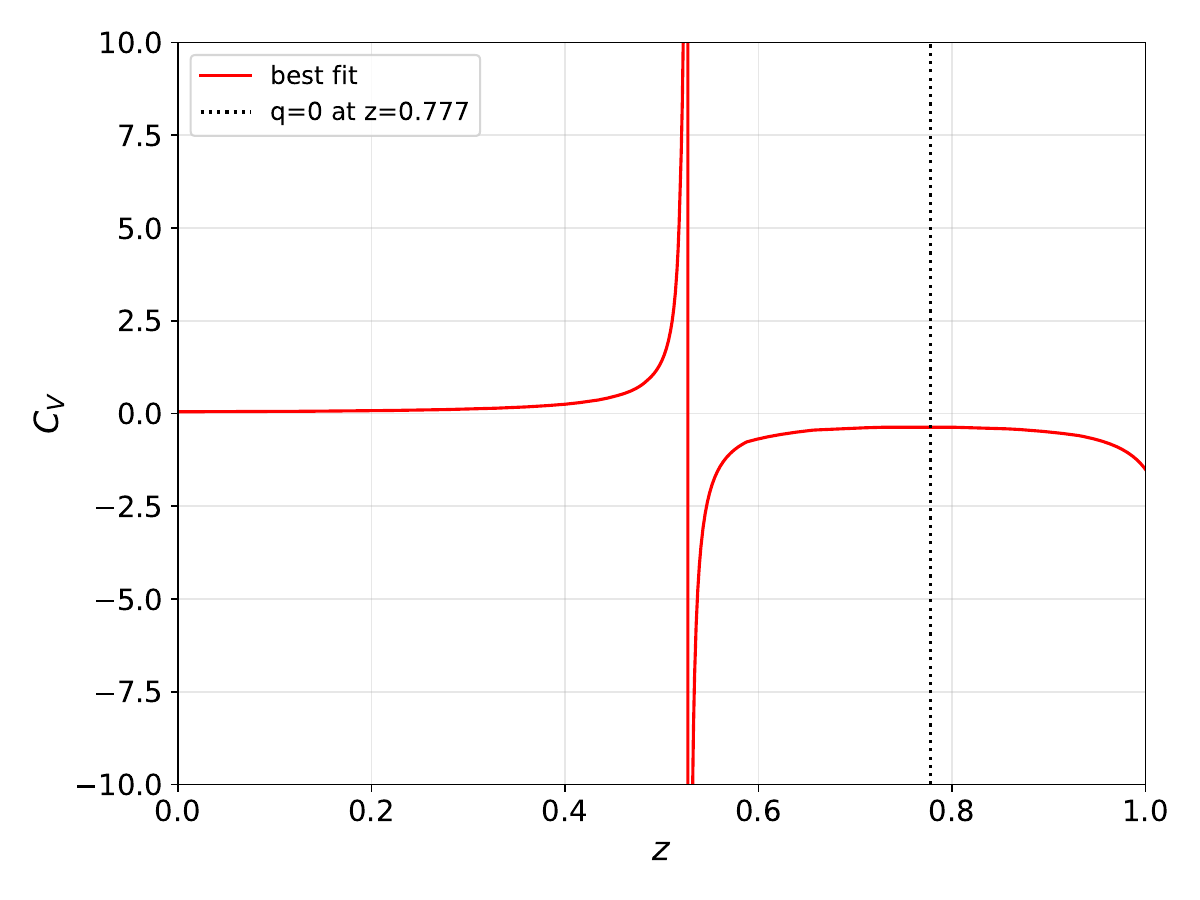}
\caption{MPL model: same as Fig.~\ref{fig:q_cv_LCDM}.}
\label{fig:q_cv_MPL}
\end{figure*}

\begin{figure*}[ht]
\centering
\includegraphics[scale=0.37]{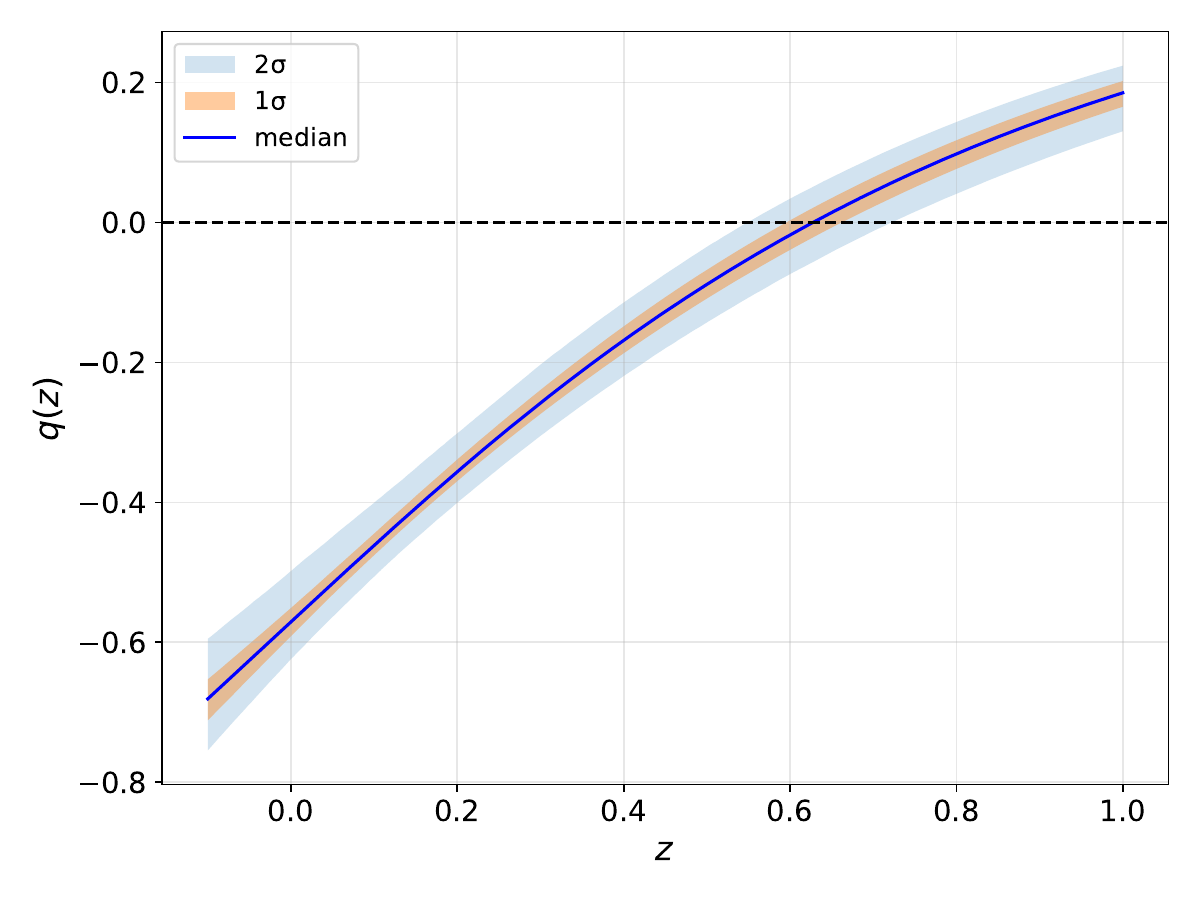}
\includegraphics[scale=0.37]{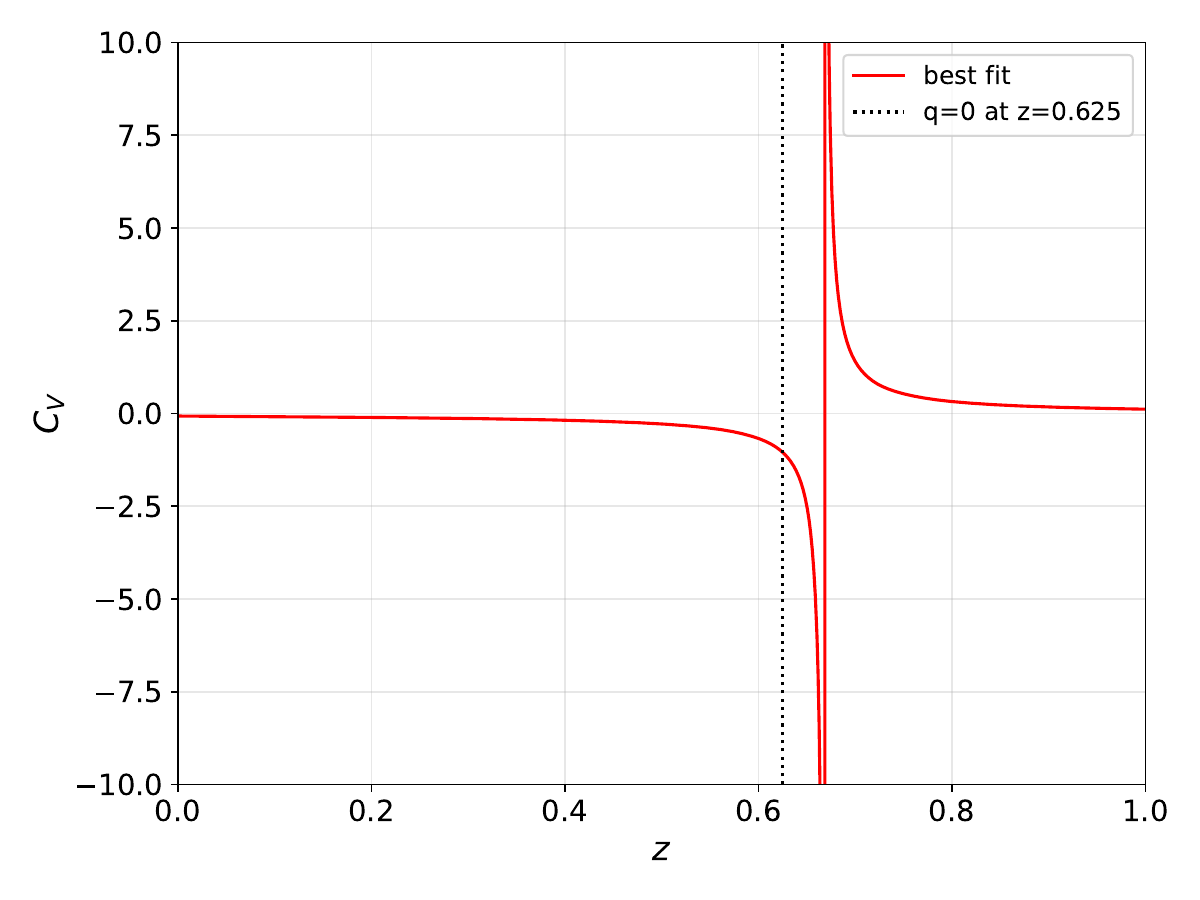}
\caption{MmAH model: same as Fig.~\ref{fig:q_cv_LCDM}.}
\label{fig:q_cv_MAH}
\end{figure*}

Figures~\ref{fig:q_cv_LCDM}-\ref{fig:q_cv_MAH} show the deceleration parameter $q(z)$ and the best-fit reconstruction of heat capacity $C_V(z)$ (right panels;the heat-capacity reconstruction is shown only for the best-fit cosmology because $C_V$ depends on the second derivative of the expansion history, $\ddot H$, and propagating the full posterior distribution significantly broadens the divergence structure and smears the pole, particularly for the dynamical dark energy models). The transition redshift $z_{q=0}$, at which $q$ changes from positive to negative, is well constrained. For $\Lambda$CDM we obtain $z_{q=0}=0.667\pm0.02$, while the dynamical models show mild shifts: $0.739\pm0.04$ for CPL, $0.777\pm0.04$ for MPL, and a slightly lower value of $0.625\pm0.03$ for MmAH.

For the bestfit $\Lambda$CDM cosmology, the heat capacity $C_V$ exhibits a sharp vertical divergence at $z_{q=0}$, reproducing the theoretical behaviour reported in Ref.~\cite{Duary:2023nnf}. In contrast, the bestfit reconstructions of the dynamical dark-energy models (CPL, MPL, and MmAH) indicate that the divergence of $C_V$ occurs at redshifts different from $z_{q=0}$. This qualitative separation is visible in the bestfit reconstructions shown in the figures.

Interestingly, the bestfit $C_V$ reconstruction of MmAH is qualitatively closer to that of $\Lambda$CDM, showing a relatively sharp transition, whereas CPL and MPL exhibit broader and more intricate structures. Unlike the $\Lambda$CDM bestfit reconstruction, however, the coincidence between the heat capacity divergence and the deceleration-acceleration transition is not recovered for the dynamical dark energy parametrizations considered here.

At the present epoch ($z=0$), the sign of $C_V$ further distinguishes the models. For $\Lambda$CDM and MmAH, $C_V(0)$ is negative, while for CPL and MPL it is positive. Since $S_{UU} = -1/(T^2 C_V)$, a negative $C_V$ implies $S_{UU} > 0$(~\ref{eq:SUU}), violating the stability condition $S_{UU} \le 0$ at present~\cite{padmanabhan1990statistical,luongo2014cosmographic}, whereas a positive $C_V$ satisfies it. This difference is reflected in the Hessian analysis presented in the next subsection.

This behaviour can be understood from the denominator of $C_V$(~\ref{eq:CVdiv}),
\begin{equation}
D(z) = 2H^2\dot{H} + H\ddot{H} - \dot{H}^2,
\end{equation}
which vanishes at the divergence redshift. For $\Lambda$CDM, the condition $D(z)=0$ coincides with $q=0$, but for general dynamical dark energy models the two conditions are independent. Therefore, within the bestfit thermodynamic reconstruction presented here, the heat capacity divergence is not generically tied to the onset of cosmic acceleration. Instead, the coincidence between the two appears to be a characteristic feature of the $\Lambda$CDM-like scenario discussed in Ref.~\cite{Duary:2023nnf}.
.

\subsection{Hessian minors and thermodynamic stability}

\begin{figure*}[ht]
\centering
\includegraphics[scale=0.5]{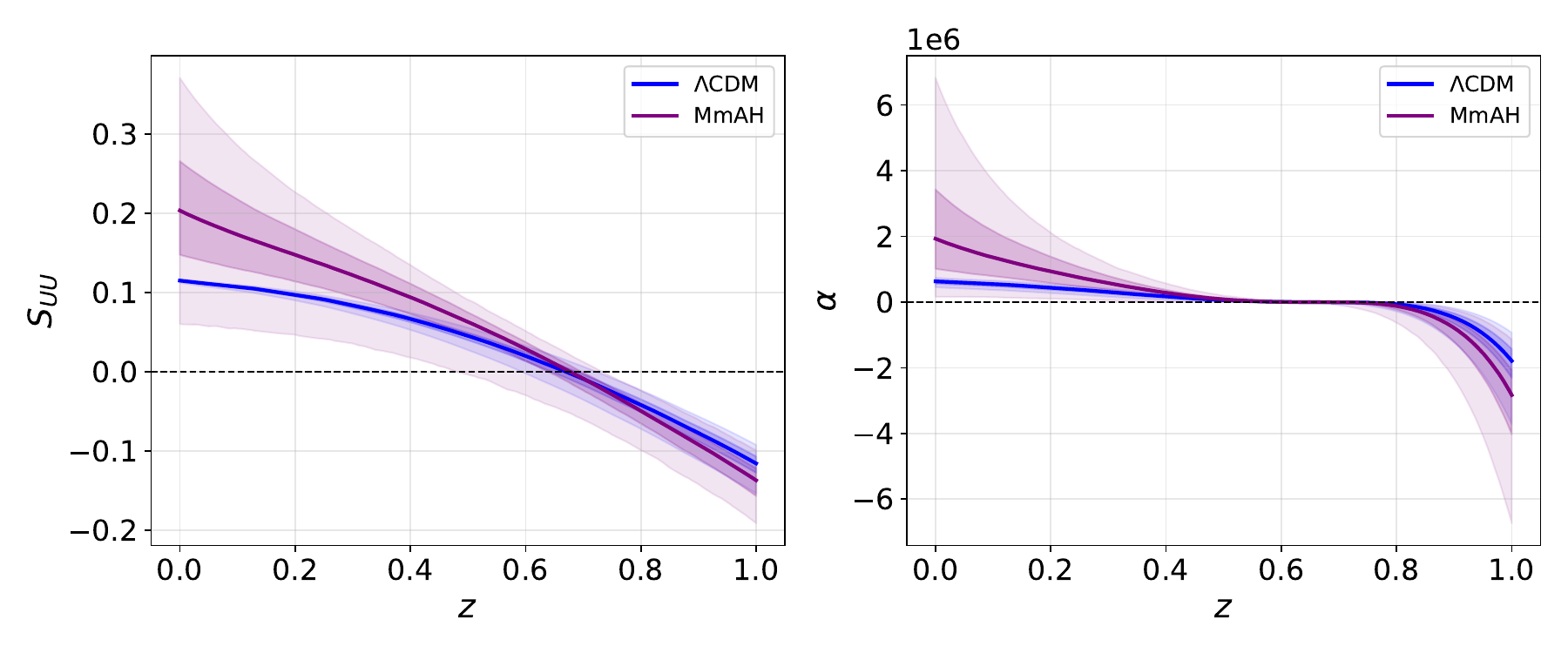}
\includegraphics[scale=0.5]{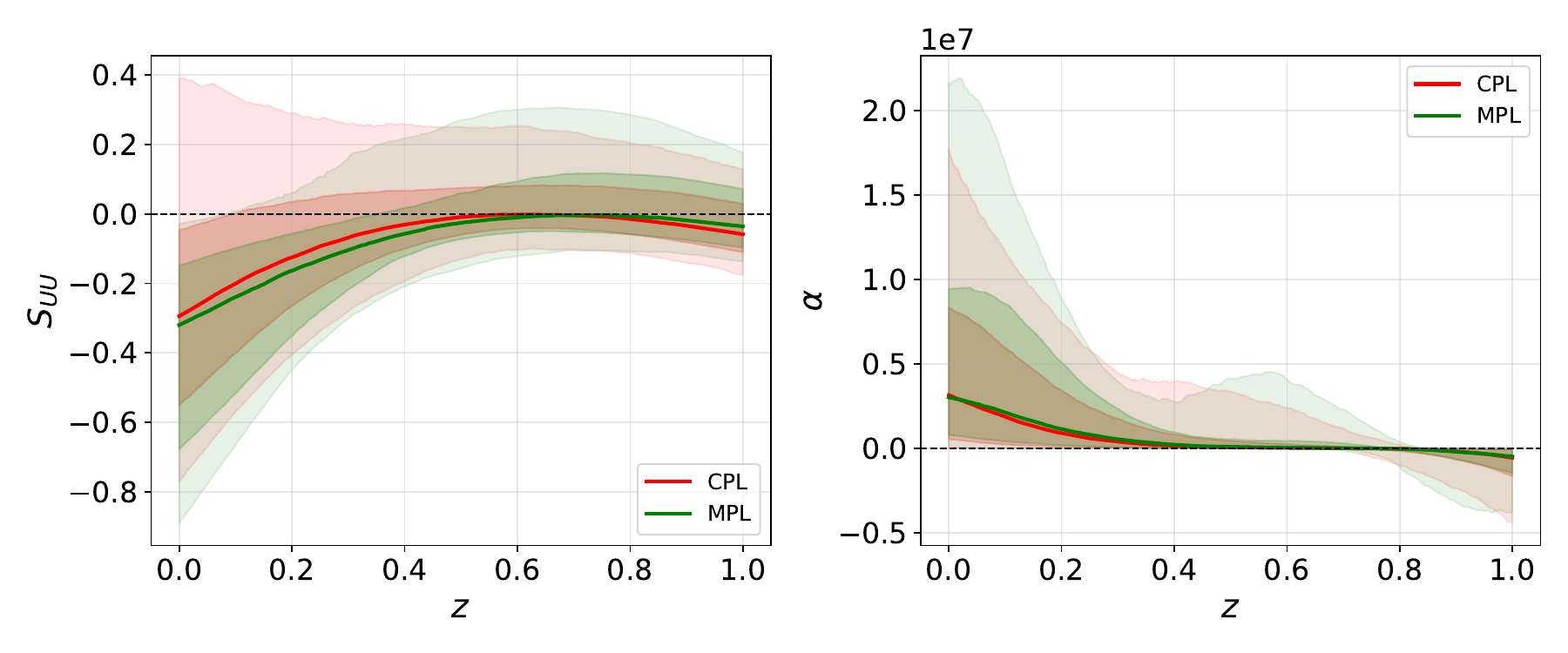}
\caption{Hessian minors: $S_{UU}$ (left) and $\alpha$ (right). Top panel: $\Lambda$CDM (blue) and MmAH (purple). Bottom panel: CPL (red) and MPL (green). Shaded regions indicate $1\sigma$ and $2\sigma$ confidence regions.}
\label{fig:hessian_comb}
\end{figure*}

Figure~\ref{fig:hessian_comb} shows the two Hessian minors that determine the local thermodynamic stability of the fluid within the horizon: $S_{UU} = \partial^2 S_{\text{in}}/\partial U^2$ and the determinant $\alpha$. According to the standard thermodynamic stability criteria~\cite{callen1998thermodynamics}, stability requires that both conditions $S_{UU} \le 0$ and $\alpha \ge 0$ be satisfied simultaneously.

For all models, the determinant $\alpha$ exhibits a sharp negative dip, indicating a transient violation of the second stability condition ($\alpha \ge 0$). This feature is associated with the thermodynamic phase transition discussed in the previous subsection. The behaviour of $S_{UU}$, however, distinguishes the models. For $\Lambda$CDM and MmAH, $S_{UU}$ becomes positive at late times, thereby violating the first stability condition ($S_{UU} \le 0$). As a result, the two stability conditions are not simultaneously satisfied at any redshift for these models.

In contrast, for CPL and MPL, $S_{UU}$ remains negative at late times ($z \lesssim 0.5$), satisfying the first condition. Consequently, both stability conditions ($S_{UU} \le 0$ and $\alpha \ge 0$) are simultaneously satisfied over a finite redshift interval, approximately $z \lesssim 0.4$, except for a brief region where $\alpha$ becomes negative. This indicates that CPL and MPL are locally thermodynamically stable in the late universe apart from a short lived instability associated with the phase transition.

Overall, while the thermodynamic phase transition induces a temporary breakdown of the second stability condition ($\alpha \ge 0$) in all models, the long term stability at low redshift is governed by the sign of $S_{UU}$, which is strongly model dependent.

\subsection{Generalized second law}

\begin{figure*}[ht]
\centering
\includegraphics[scale=0.37]{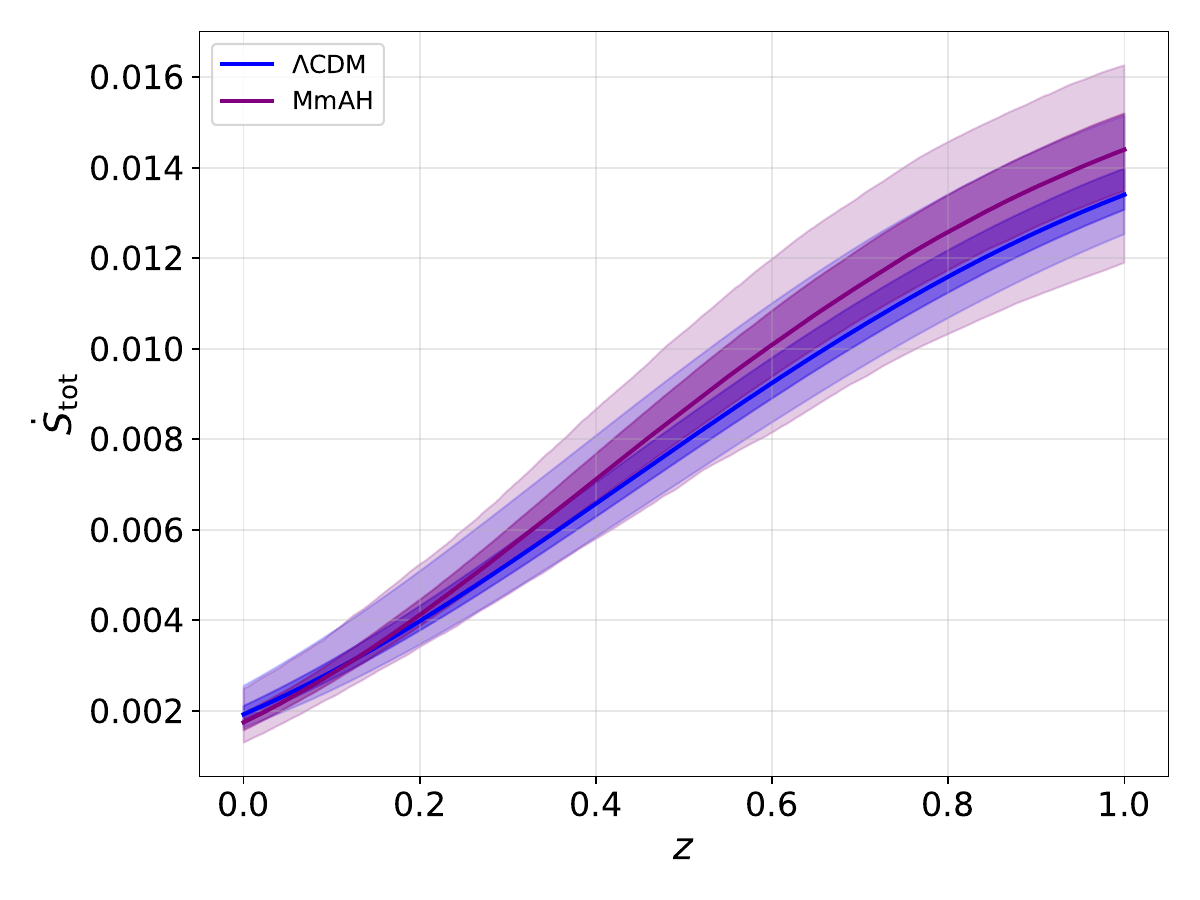}
\includegraphics[scale=0.37]{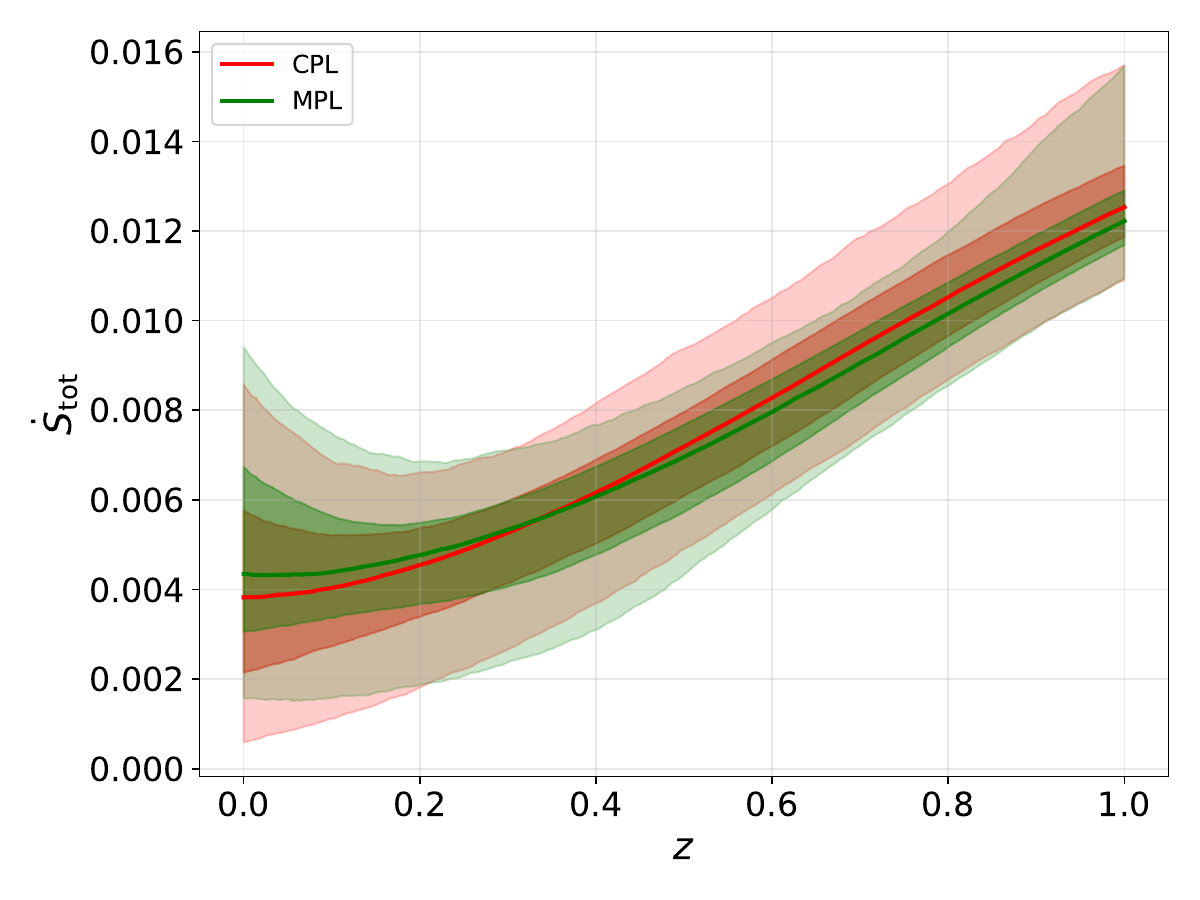}
\caption{Entropy production rate $\dot{S}_{\text{tot}}(z)$. Top panel: $\Lambda$CDM (blue) and MmAH (purple). Bottom panel: CPL (red) and MPL (green). Shaded regions represent the $1\sigma$ and $2\sigma$ confidence intervals.}
\label{fig:gsl_comb}
\end{figure*}

Figure~\ref{fig:gsl_comb} presents the total entropy production rate $\dot{S}_{\text{tot}}(z)$ computed using the Hayward-Kodama temperature. For all four models, the median value of $\dot{S}_{\text{tot}}(z)$ remains strictly positive throughout the redshift interval $0 \le z \le 1$, while the corresponding $1\sigma$ and $2\sigma$ confidence regions lie entirely above zero. This provides strong evidence for the validity of the generalized second law (GSL) in all cosmological models considered in this work.

Across the full redshift range investigated, no statistically significant violation of the condition $\dot{S}_{\text{tot}} \geq 0$ is observed. The probability that the entropy production rate remains positive is therefore consistent with unity for all models.

To understand the origin of this robust positivity even in scenarios where the dark energy equation of state crosses the phantom divide, we examine the factor $2H^2+\dot{H}$ appearing in the denominator of $\dot{S}_{\text{tot}}$. Using the Friedmann relation
\begin{equation}
\dot{H}=-\frac{3}{2}H^2(1+w_{\rm eff}),
\end{equation}
where
\begin{equation}
w_{\rm eff} \equiv \frac{p_{\rm tot}}{\rho_{\rm tot}}
\end{equation}
denotes the effective equation of state of the total cosmic fluid, we obtain
\begin{equation}
2H^2+\dot{H}
=\frac{H^2}{2}\left(1-3w_{\rm eff}\right).
\end{equation}
Since all remaining factors entering $\dot{S}_{\text{tot}}$ are positive, the sign of the entropy production rate is determined by the quantity $1-3w_{\rm eff}$. Consequently, the generalized second law is satisfied whenever the effective equation of state obeys $w_{\rm eff}<1/3$.

For all observationally constrained models considered here, we find that $w_{\rm eff}(z)<1/3$ throughout the interval $0\le z\le1$. As a result, $2H^2+\dot H$ remains positive and hence $\dot S_{\rm tot}>0$ over the entire redshift range. Therefore, the generalized second law is naturally satisfied in all models studied in this work, in contrast to the violations reported in Ref.~\cite{Duary:2019dfu}.

\section{Conclusion}
\label{sec:con}

In this work, we have carried out a comprehensive observational and thermodynamic analysis of four cosmological models—$\Lambda$CDM, CPL, MPL, and the three parameter MmAH parametrization—using a full likelihood framework. While the MPL model has been previously studied and the MmAH model was originally proposed without a complete analysis including the full CMB likelihood, here we perform a consistent and unified parameter estimation for all models. This enables a direct and fair comparison of their performance under current observational constraints. The parameter estimation is performed using the nested sampling algorithm PolyChord, which is particularly well suited for exploring degenerate and multi-dimensional parameter spaces. This is especially important for the dynamical dark energy models considered here, where parameter degeneracies can significantly affect the inference. From the statistical analysis, all dynamical dark energy models provide an improved fit relative to $\Lambda$CDM, with a reduction in $\chi^2$ of $\Delta\chi^2 \sim 6$-$6.5$. Among them, the MPL parametrization yields the best overall fit, indicating that its functional form captures features of the late-time expansion history beyond the standard cosmological constant. The MmAH model, despite having three free parameters, achieves a comparable improvement in $\chi^2$, demonstrating its viability as a dynamical dark energy candidate. However, one of its parameters remains weakly constrained, suggesting that current data are not yet sufficient to fully resolve its parameter space. Although the dynamical dark-energy models provide a noticeably better fit to the data in terms of $\chi^2$, the Bayesian evidence does not indicate a statistically significant preference over $\Lambda$CDM. This suggests that the improved fit is largely balanced by the increased model complexity, highlighting the importance of considering both goodness-of-fit and model selection criteria when assessing extensions of the standard cosmological model.

A key novel aspect of this work is the reconstruction of thermodynamic quantities directly from observationally constrained models.
Since these quantities depend on the expansion history and its higher derivatives, they should be interpreted within the framework of observationally constrained cosmological models rather than as direct model independent reconstructions from the data. Extending previous purely theoretical studies—most notably Ref.~\cite{Duary:2023nnf}, which focused only on $\Lambda$CDM without observational input—we provide the first data driven thermodynamic analysis across multiple dark energy parametrizations. This allows for a robust assessment of thermodynamic behaviour in realistic cosmological scenarios.Among the reconstructed thermodynamic quantities, the heat capacity $C_V$ is particularly sensitive to the second derivative of the expansion history. Although a full posterior reconstruction of $C_V$ would, in principle, provide a more complete statistical characterization, its explicit dependence on $\ddot{H}$ causes the propagated posterior uncertainties to significantly broaden the divergence structure, smear the pole, and, for the dynamical dark energy parametrizations, even produce multiple apparent divergences. Consequently, a statistically robust determination of the divergence redshift from the posterior samples is not feasible. For this reason, we present the bestfit reconstruction of $C_V$ to illustrate the thermodynamic behaviour implied by each observationally constrained model. The bestfit reconstructions indicate that the heat capacity divergence coincides with the deceleration-acceleration transition only for the $\Lambda$CDM model, reproducing the theoretical behaviour reported in Ref.~\cite{Duary:2023nnf}. In contrast, the bestfit reconstructions of the dynamical dark energy parametrizations (CPL, MPL, and MmAH) exhibit qualitatively different behaviour, with the heat capacity divergence occurring at redshifts different from $z_{q=0}$. Therefore, within the framework of the bestfit thermodynamic reconstruction, these results suggest that the correspondence between the heat capacity divergence and the onset of cosmic acceleration is not a universal feature but depends on the adopted dark energy parametrization. The generalized second law is satisfied for all models over the redshift range $0 \le z \le 1$, with $\dot{S}_{\text{tot}} > 0$ throughout. This behaviour follows from the condition $w_{eff}(z) < 1/3$, which is respected by all observationally viable models considered here, including those crossing the phantom divide. Thus, the generalized second law emerges as a robust and model independent consistency condition in the late universe. The Hessian stability analysis further distinguishes the models. While all models exhibit a transient violation of the second stability condition ($\alpha \ge 0$) at the phase transition, the long term behaviour depends on $S_{UU}$. The CPL and MPL models satisfy both stability conditions over a finite late time interval, whereas $\Lambda$CDM and MmAH fail to do so due to $S_{UU} > 0$ at low redshift. This indicates that thermodynamic stability can provide an additional discriminator between competing dark energy models.

In summary, our results demonstrate that thermodynamic properties reconstructed within observationally constrained cosmological models offer valuable insights into the nature of dark energy. While all dynamical dark energy parametrizations provide an improved fit to the observational data relative to $\Lambda$CDM in terms of $\chi^2$, the Bayesian evidence shows a mild preference for the $\Lambda$CDM model due to its lower complexity. The bestfit thermodynamic reconstruction indicates that the coincidence between the heat-capacity divergence and the onset of cosmic acceleration is recovered only for the $\Lambda$CDM model, whereas the dynamical dark energy parametrizations exhibit qualitatively different thermodynamic behaviour. Meanwhile, the generalized second law remains universally valid across all viable models considered in this work.

Future surveys such as Euclid, the Nancy Grace Roman Space Telescope, and upcoming DESI releases will significantly improve constraints on the dark energy equation of state. In particular, these observations will be crucial for testing models like MmAH, where additional parameters are currently weakly constrained, and for determining whether the thermodynamic behaviour identified in this work persists with higher precision data.

\section*{Acknowledgments}

The author acknowledges the High Performance Computing facility \textit{Pegasus} at IUCAA, Pune, India, for providing the computational resources used in this work. Sonej Alam acknowledges Prof. Somasri Sen, Dr.~Wali Hossain and Tanima Duary for insightful discussions. He also gratefully acknowledges financial support from the Council of Scientific and Industrial Research (CSIR), Government of India, through the CSIR NET-SRF fellowship (File No.~09/0466(12904)/2021).

\bibliographystyle{elsarticle-num}
\bibliography{ref}

\end{document}